# A Dual Field-of-View Zoom Metalens

Guoxing Zheng[1, 5*], Weibiao Wu[1], Zile Li[1, 5], Shuang Zhang[2*], Muhammad Qasim Mehmood[3,4], Ping'an He[1] and Song Li[1, 5]

1. Electronic Information School, Wuhan University, Wuhan, 430072, China
2. School of Physics & Astronomy, University of Birmingham, Birmingham, B15 2TT, UK
3. Information Technology University of the Punjab, Lahore 54000, Pakistan
4. Department of Electrical and Computer Engineering, National University of Singapore, 4 Engineering Drive 3, Singapore, Singapore 117583
5. Cooperative Innovation Center of Geospatial Technology, Wuhan University, Wuhan, 430079, China

Email: gxzheng@whu.edu.cn, s.zhang@bham.ac.uk

**Abstract:** Conventional optical zoom system is bulky, expensive and complicated for real time adjustment. Recent progress in the metasurface research has provided a new solution to achieve innovative compact optical systems. In this paper, we propose a highly integrated zoom lens with dual field-of-view (FOV) based on double sided metasurfaces. With silicon nanobrick arrays of spatially varying orientations sitting on both side of a transparent substrate, this ultrathin zoom metalens can be designed to focus an incident circular polarized beam with spin-dependent FOVs without varying the focal plane, which is important for practical applications. The proposed dual FOV zoom metalens, with the advantages such as ultracompactness, flexibility and replicability, can find applications in fields which require ultracompact zoom imaging and beam focusing.

Zoom lens system is one of the most important optical systems and its applications can be found in various imaging systems. In a widely-used mechanical zoom lens, to keep the image staying in focus throughout the zoom range, one needs a movable component providing the change in magnification and a shift of another component to compensate the defocus [1]. Compared with continuous zoom lens systems, a dual or multiple field-of-view lens system has relatively low structure complication and costs. The applications of such lenses include vigilance [1,2], detection [3], remote sensing [4,5] and laser processing [4,6]. For example, in target-tracking application, a zoom optical system is required to search the target in a wide field-of-view and then switch to a narrow field-of-view for tracking and identification while maintaining the focal plane unchanged [7,8]. When it comes to the laser processing fields, zoom lens can be used to focus an incident laser beam into spots of different sizes at different working conditions. In most traditional dual field-of-view optical system designs, the optical configurations are often based on the mechanical motion of the lens groups or liquid crystal lenses [9-12], which increase the manufacture costs and bring inconvenience to practical applications.

With the ability to manipulate the electromagnetic fields, two-dimensional (2D) planar metamaterials, or so called metasurfaces, have become a topical research area within the field of nanophotonics [13-18]. By virtue of their capabilities for complete control over the phase of the optical fields, metasurfaces can operate as high-performance optical devices, such as a flat metalens and holograms. Since a metalens based on geometric phase is sensitive to the polarization state of an incident light, it may enable the design of the dual field-of-view optical system. In this paper, we propose a novel composite metalens to realize reconfigurable optical zooming through controlling the polarization states of an incident beam. With metasurfaces fabricated on both sides of a glass substrate, the zoom metalens does not require any axial motion that traditional counterparts always entail. Further, because of its ultracompact dimensions, metasurface zoom lens enables the miniaturization of conventional refractive optics into planar structures and can be potentially integrated with the focal-plane arrays (FPAs).

Among different types of metasurfaces, we choose dielectric nanobrick arrays operating as a transmission-type geometric metasurface to control the phase and polarization of an incident beam [18,19]. These nanobricks, sitting on a silicon dioxide substrate and having the same cell dimensions but different orientations angles, can transform an incident circularly polarized light into the opposite polarization state with high efficiency. The flip of spin introduces an accurate phase delay which is twice the orientation angle [18-24]. These nanobrick arrays can act as a continuous-phase modulated lens, but only requiring structures of two-step depth, which holds promise for designing dual field-of-view optical system with high-performance and low costs.

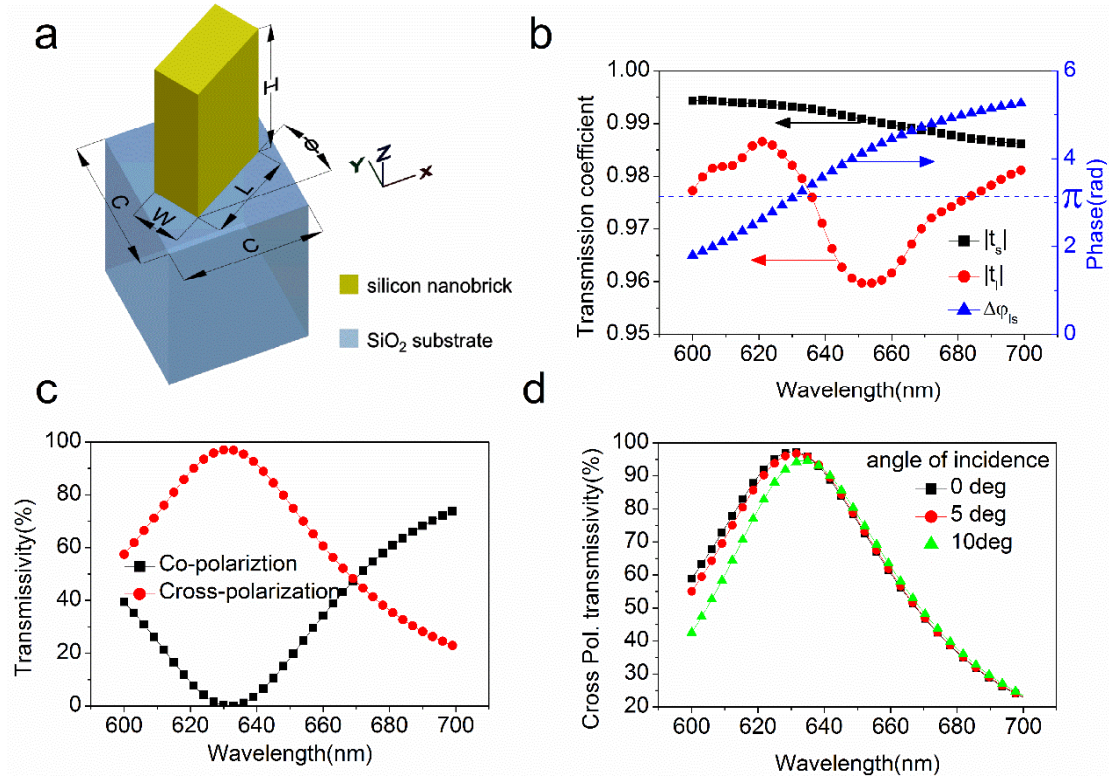

**Figure 1| Illustration of the unit-cell structure and its polarization conversion efficiency by numerical simulations. a,** Schematic diagram of one nanobrick cell. A nanobrick is optimally designed and fixed with cell size C = 300 nm, length L = 140 nm, width W = 70 nm, and height H = 350 nm. The operation wavelength is 635 nm. **b,** Simulated transmission coefficients $|t_l|$, $|t_s|$ and phase difference $\Delta\phi_{ls}$ versus wavelength, where l and s denote the long and short axis directions of the nanobricks, respectively. **c,** Simulated cross-polarization and co-polarization transmissivity with normal light incidence. **d,** Simulated cross-polarization transmissivity versus wavelength for different incident angles.

A schematic diagram of a single nanobrick unit cell is shown in Fig. 1a. The corresponding transmission coefficients for electric field aligned along the long and short axis of the nanobrick is presented in Fig. 1b. One can see that the phase difference between the transmission coefficients $t_l$ and $t_s$ is approximately $\pi$ in the wavelength range of 630-640 nm; moreover, such configuration maintains very large transmission coefficient over 0.96 for both linear polarizations. Therefore, each nanobrick functions as a sub-wavelength half waveplate [18-22], leading to nearly unity conversion between the LCP and RCP at 635 nm wavelength, as shown in Fig. 1c. This circular polarization conversion is accompanied by a geometric phase delay $2\phi$ ($\phi$ is the orientation angle of the nanobrick). Therefore, only by changing the orientation angles of each nanobrick, one can control the wavefront of an incident beam at the scale of each individual unit cell. The transmission spectra for three different incident angles are shown in Fig. 1d. It shows that the transmission of a nanobrick maintains a large value (over 90%) at different incident angles (up to 10°) at a wavelength around 635 nm, which is promising for designing high performance zoom metalens.

Since the geometric metalens is sensitive to the handedness of the incident CP light, a positive metalens can be converted to a negative lens (vice versa) if the handedness of light is reversed. Therefore, by cascading two metalenses of the same polarity under the illumination of the same CP, the second metalens reverses its polarity due to the flip of the handedness of light after transmitting through the first one. In this way, one can obtain two different combinations of focal lengths of the two lenses for incident beam of different spins. This forms the basic principle of a dual FOV zoom metalens, which is schematically illustrated by Fig. 2, where the combination of polarizer and QWP generates the desired circular polarization state before light entering the composite metalens system.

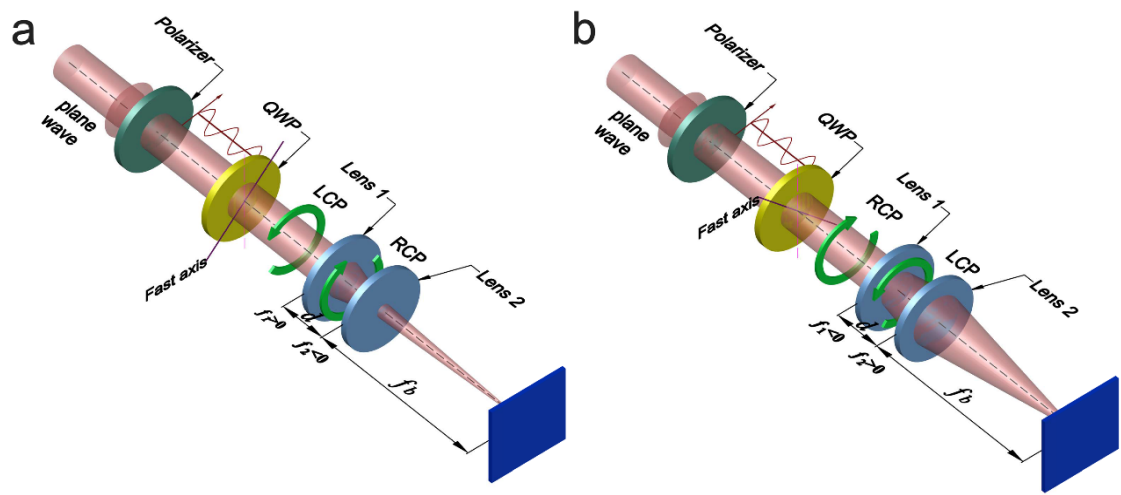

**Figure 2 | Layout and illustration of the zoom mechanism of metalens. a**, a narrow FOV mode characterized by a LCP incident light. **b**, a wide FOV mode characterized by a RCP incident light.

We next deduce the basic parameters of a composite lens illustrated in Fig. 2a. It is known that the focal length of a composite lens can be expressed as

$$F_1' = \frac{f_1' f_2'}{f_1' + f_2' - d} \quad (1)$$

where $f_1'$ and $f_2'$ are the focal lengths of lens 1 and lens 2, respectively, $d$ is the distance between the two lenses. The distance between the focal plane and the back plane of lens 2, or so called the back focal length can be expressed as

$$f_{b1} = \frac{f_2'(f_1' - d)}{f_1' + f_2' - d}. \quad (2)$$

By rotating the QWP by 90 degrees (Fig. 2b), the CP of incident light is transformed into its opposite CP state. Accordingly, the polarities of the two metalenses are reversed simultaneously, resulting in focal length of $-f_1'$ and $-f_2'$ respectively. Hence the combined focal length and the back focal length of the composite lens is changed to

$$F_2' = \frac{-f_1' \cdot (-f_2')}{-f_1' - f_2' - d} = \frac{f_1' f_2'}{-f_1' - f_2' - d} \quad (3)$$

and

$$f_{b2} = \frac{-f_2'(-f_1' - d)}{-f_1' - f_2' - d} = \frac{f_2'(f_1' + d)}{-f_1' - f_2' - d}. \quad (4)$$

For practical applications, it is desired that the image plane of a zoom lens remains unchanged. By setting $f_{b1} = f_{b2}$ and with their expressions given by Eqn. (2) and (4), we get

$$f_1'(f_1' + f_2') = d^2. \quad (5)$$

As an important parameter of a dual field-of-view optical system, the zoom ratio between the narrow and the wide FOV modes of the system must be considered in our analysis. Here, if we assume $f_1' > 0$ and $f_2' < 0$, by combining equations (1), (3), and (5), we can obtain the condition $F_1' > F_2'$. Hence, by defining $F_1' = \beta F_2'$, where $\beta$ ($\beta > 1$) is the optical zoom ratio of the composite lens, we can obtain another expression when combing equation (1) and (3):

$$d = \frac{\beta + 1}{\beta - 1}(f_1' + f_2'). \quad (6)$$

Eqns (5) and (6) formulate the constraints among $f_1'$, $f_2'$ and $d$, which can be easily satisfied with the metalens configuration proposed here.

To achieve the phase profile, we arrange two metalenses on both sides of a transparent substrate, which forms a highly integrated zoom metalens, as shown in Fig. 3. The following expression governs the relationship between the rotation angle $\phi$ and the location of each nanobrick for the front lens (Lens 1):

$$\phi_1(r_1) = \pm \frac{\pi}{\lambda/n}\left[\sqrt{r_1^2 + (nf_1')^2} - |nf_1'|\right] \quad (7)$$

where $\lambda$ is the free-space wavelength, $r_1$ is the distance that each nanobrick departures the center of Lens 1, $f_1'$ is the focal length and $n$ is the refractive index of the silica substrate.

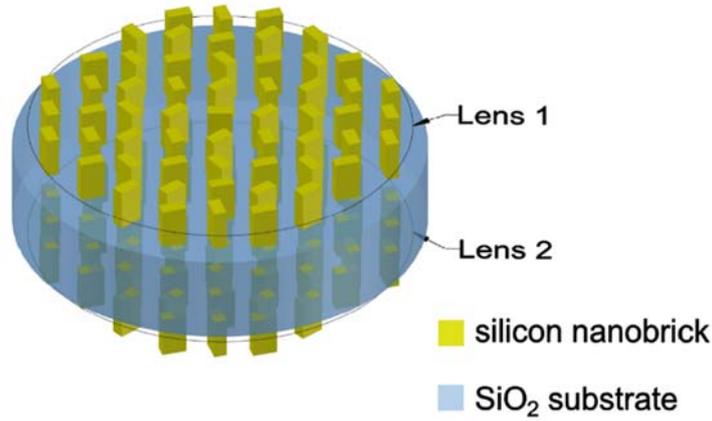

**Figure 3 | Schematic diagram of a highly integrated zoom metalens with dual FOVs.** The two metalenses are arranged on both side of a silicon dioxide transparent substrate.

For the rear lens (Lens 2), the propagation of refractive light from Len 1 is different for the two FOV working modes, as shown in Fig. 4. For their overlapping area at the surface of Len 2 which satisfies $r_2 \leqslant S$ (S is the edge height for the long focal length mode and is given by $\frac{D}{2|nf_1'|}(|nf_1'| - d)$), we only consider the condition for optimizing the focusing of the long focal length mode and the rotation angle of each nanobrick satisfies

$$\phi_2(r_2) = \pm \frac{\pi}{\lambda}\left[n\sqrt{(|nf_1'| - d)^2 + r_2^2} - n(|nf_1'| - d) + f_b - \sqrt{r_2^2 + f_b^2}\right]. \quad (8)$$

Despite that this phase distribution at the overlapping area cannot operate perfectly for the short focal length mode due to the deviation from the paraxial approximation, the two different beams can converge reasonably well for a small numerical aperture at these areas.

Next, if $r_2 > S$, then the rotation angle is determined only by short focal length mode:

$$\phi_2(r_2) = \pm \frac{\pi}{\lambda} \left[ n\sqrt{(|nf_1'| + d)^2 + r_2^2} - n(|nf_1'| + d) - f_b + \sqrt{r_2^2 + f_b^2} \right]. \quad (9)$$

To investigate the performance of our proposed double-side zoom metalens, ray tracing was performed to determine the location of the incident rays. It is known that the refracted light of geometrical metasurfaces is governed by [13]

$$\sin(\theta_t) n_t - \sin(\theta_i) n_i = \frac{2}{k_0} \cdot \frac{d\phi}{dr}, \quad (10)$$

where $\theta_i$ and $\theta_t$ are the incident and refracted angles, $n_i$ and $n_t$ are the refractive indices in the first and second media, respectively, $k_0$ is the wavenumber in vacuum. Combining equations (7) to (10), one can derive that

$$\begin{cases} \theta_{t_1} = \sin^{-1}\left(\frac{2}{k_0 n} \cdot \frac{d\phi_1(r_1)}{dr_1} + \frac{\sin \theta_{i1}}{n}\right) \\ r_2 = r_1 - d \cdot \tan \theta_{t_1} \\ \theta_{t_2} = \sin^{-1}\left(\frac{2}{k_0} \cdot \frac{d\phi_2(r_2)}{dr_2} + \sin \theta_{i2} \cdot n\right) \\ r_3 = r_2 - f_b \tan \theta_{t_2} \end{cases}. \quad (11)$$

The ray tracing results based on equation (11) are shown in Fig. 4. One can clearly observe two beams (red and green) with different circular polarizations are focused at the same focal plane, but with dramatically different numerical apertures that correspond to different FOVs.

The dual FOV zoom metalens presented in Fig. 4 is designed with the following parameters to satisfy Eqns (5) and (6): the two metalenses have focal lengths of $f_1' = 3/2$ mm, $f_2' = -5/6$ mm for a LCP incident light, and the thickness of the transparent glass substrate equals to 1.457 mm. The designed dual FOV zoom metalens has a zoom ratio of 5 and the composite focal lengths are 3.75 mm and 0.75 mm, respectively. For both incident circular polarizations, the final rays focus at the same position located at the theoretically focal plane ($f_b$=1.25 mm).

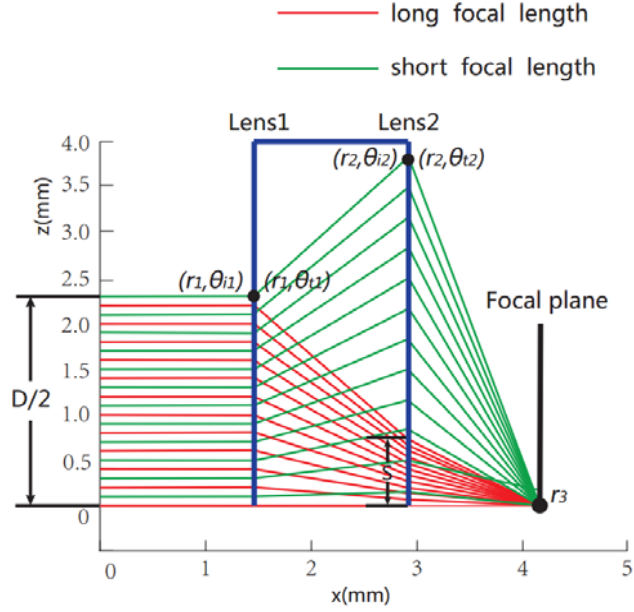

**Figure 4 | Ray tracing of zoom metalens with two different working conditions.** Red and green light rays are for long and short focal length modes, respectively.

Further numerical validation is carried out by a full wave simulation using a commercial software package (Comsol) to simulate the propagation of a CP light through the dual FOV zoom metalens. To facilitate the simulation, all the geometric parameters of the zoom metalens are reduced to one thousandth of their original values (e. g., $f_1'$ shrinks from 3/2 mm to 3/2 μm, $f_2'$ shrinks from -5/6 mm to -5/6 μm and $d$ shrinks from 1.457 mm to 1.457 μm). Consequently, only 21 silicon nano-antennas are arranged on both sides of a glass substrate along the $x$ direction such that each metalens function as a cylindrical lens. The metalens operates at a visible wavelength of 635 nm. The intensity distribution of the electric field for an incident CP light with different handedness is shown in Figs. 5a and 5b, respectively. It is observed that for two different working modes of the zoom metalens, the focus planes almost coincide with each other ($f_b \approx 1.26$ μm), which agrees well with our theoretical value of 1.25 μm according to equation (2) or (4). One further observes that the dimensions of the focus spots are different for the two working modes. Since any lens is diffraction-limited and the resolution is governed by

$$\text{Res} = 0.61 \frac{\lambda}{NA}, \tag{12}$$

where *NA* is the numerical aperture of the metalens. According to equation (12), for the same lens aperture here, a metalens with longer focal length corresponds to a smaller *NA* and would produce a larger spot, which is consistent with Figs. 5a and 5b.

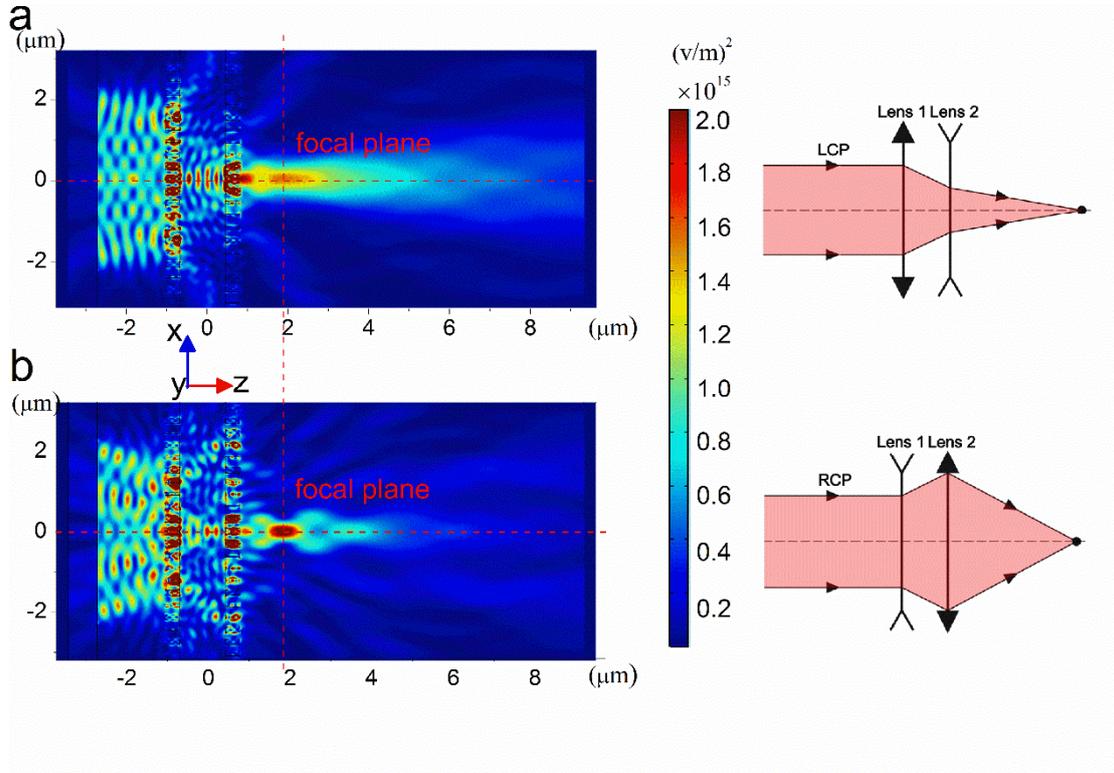

**Figure 5 | Full-wave Numerical simulations of a dual FOV zoom metalens with a normal incident CP light. a,b**, Simulated intensity distribution with a LCP (**a**) and RCP (**b**) normally incident light. For the same nanostructures, zoom metalens with long focal length (right upper) and short focal length (right below) are realized only by changing the handedness of the incident light.

To further investigate the characteristics of the dual FOV zoom metalens, we simulate the propagation of light through the designed metalens for two different cases: a plane wave with an oblique incident angle, and a point source. The intensity distribution is shown in Fig. 6a for a RCP plane wave at an incident angle of 2°. It shows that the transmitted beam is strongly focused at a position of $x \approx 0.03$ μm and $z \approx 1.96$ μm, which agrees well with our theoretical calculations $x = F_2' \cdot \tan 2° = 0.026$ μm. In Fig. 6b, the intensity distribution is shown when a RCP point light source is placed at a position of $x = 0.3$ μm and $z = -2$ μm. The image point appears at a position of $x \approx -0.18$ μm and $z \approx 2.57$ μm, which is again almost the same as the theoretical results.

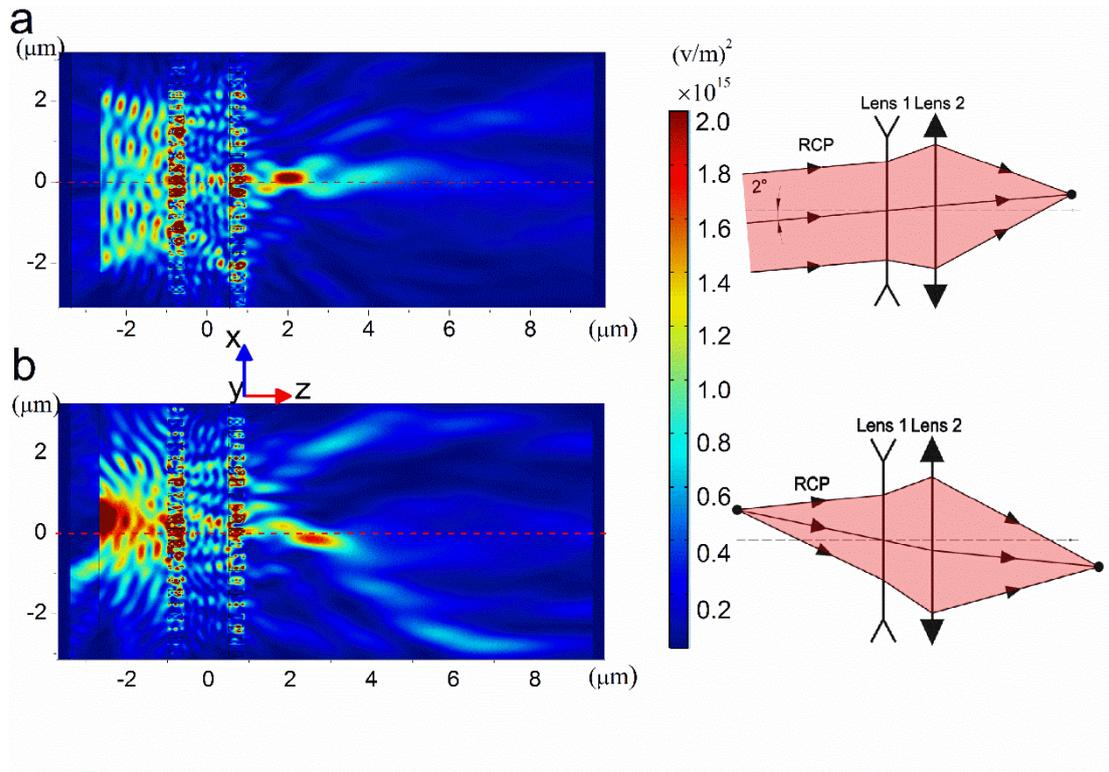

**Figure 6 | Full-wave Numerical simulations of a dual FOV zoom metalens with different imaging situations.**

**a**, Intensity distribution with a RCP plane wave (slant angle 2°) propagating through the dual FOV zoom metalens.

**b**, Intensity distribution with a RCP point light source located at a position of x= 0.3 μm and z=-2 μm propagating through the dual FOV zoom metalens. For the same nanostructures, the layout of the zoom metalenses illuminated with slanted plane wave (right upper) and point source (right below) is illustrated.

In summary, we have proposed a zoom metalens whose FOV can be simply switched by flipping the helicity of incident light, while the focal plane remain unchanged. Compared with conventional zoom lens, the dual FOV zoom metalens based on polarization control has important advantages such as simplicity, ultra-compactness, flexibility and replicability. The zoom metalens can be fabricated in low costs and it is promising to be integrated into cell phones and many others imaging system to form a highly integrated optical zoom camera lens.